\documentclass[twocolumn]{revtex4}
\usepackage{graphicx}
\usepackage{dcolumn}
\usepackage{natbib}

\newcommand{\units}[1]{\ensuremath{\mathrm{#1}}}
\newcommand{\amount}[2]{\ensuremath{#1\:\units{#2}}}

\begin{document}

\title{Fast tunnel rates in Si/SiGe one-electron single and double
  quantum dots}

\author{Madhu Thalakulam}
\affiliation{University of Wisconsin-Madison, Madison, Wisconsin 53706, USA}
\author{C. B. Simmons}
\affiliation{University of Wisconsin-Madison, Madison, Wisconsin 53706, USA}
\author{B. M. Rosemeyer}
\affiliation{University of Wisconsin-Madison, Madison, Wisconsin 53706, USA}
\author{D. E. Savage}
\affiliation{University of Wisconsin-Madison, Madison, Wisconsin 53706, USA}
\author{M. G. Lagally}
\affiliation{University of Wisconsin-Madison, Madison, Wisconsin 53706, USA}
\author{Mark Friesen}
\affiliation{University of Wisconsin-Madison, Madison, Wisconsin 53706, USA}
\author{S. N. Coppersmith}
\affiliation{University of Wisconsin-Madison, Madison, Wisconsin 53706, USA}
\author{M. A. Eriksson}
\affiliation{University of Wisconsin-Madison, Madison, Wisconsin 53706, USA}

\begin{abstract}
  We report the fabrication and measurement of one-electron single and
  double quantum dots with fast tunnel rates in a Si/SiGe
  heterostructure.  Achieving fast tunnel rates in few-electron dots
  can be challenging, in part due to the large electron effective mass
  in Si. Using charge sensing, we identify signatures of tunnel rates
  in and out of the dot that are fast or slow compared to the
  measurement rate.  Such signatures provide a means to calibrate the
  absolute electron number and verify single electron
  occupation. Pulsed gate voltage measurements are used to validate
  the approach.
\end{abstract}

\maketitle

The spins of electrons isolated in quantum dots are promising
candidates for solid-state qubits.\cite{Loss:1998p120} Spin readout
and manipulation have been demonstrated in GaAs quantum dots, using
both one and two-electron spin states as logical
qubits.\cite{Elzerman:2004p431,Petta:2005p2180,Koppens:2006p766} To
enable successful error correction, it is advantageous to have spin
dephasing times and lifetimes as long as possible.  Because isotopes
of Si with zero nuclear spin exist, resulting in particularly slow
electron spin dephasing and long
lifetimes,\cite{Tahan:2002p035314,Tyryshkin:2003p193207,deSousa:2003p1660}
a number of proposals have been made to construct spin qubits based on
confined electrons in Si quantum dots and
donors.\cite{Kane:1998p133,Vrijen:2000p1643,Friesen:2003p121301,Hollenberg:2006p1534,Morello:2009p081307}
There has been significant progress in the development of quantum dots
in
Si,\cite{Slinker:2005p246,Berer:2006p162112,Angus:2007p845,Hu:2007p465,Zimmerman:2007p033507,Shaji:2008p540,Liu:2008p073310,Lansbergen:2008p1545,Fuhrer:2009p707,Simmons:2009p3234,Nordberg:2009p115331,Hayes:2009preprint,Xiao:2009preprint}
but achieving single-charge occupation in Si dots is challenging. One
issue is the relatively large electron effective mass $m^*$ in Si,
which leads to smaller tunnel rates than would arise for the same size
and shape barrier in materials with lighter m$^*$.  While occupation
of Si single quantum dots by individual charges has been demonstrated
for both electron \cite{Simmons:2007p213103,Lim:2009p242102} and hole
\cite{Zwanenburg:2009p1071} dots, a double quantum dot with a single
electron in each dot, the foundation for coupled qubits, or the
two-electron singlet-triplet qubit,\cite{Taylor:2005p482} has not been
achieved until now.

This letter reports the demonstration of Si/SiGe single and double
quantum dots occupied by zero, one, and two electrons, and it
describes signatures of fast tunnel rates in the one-electron limit
for these dots. We show that tunnel rates in Si/SiGe quantum dots
change noticeably over moderate gate voltage ranges that correspond to
removing several electrons from the quantum dot.  This change is rapid
enough to require retuning of the tunnel barriers in and out of the
dot when approaching the one-electron state. By careful tuning of the
gate voltages, we can measure reliably the expulsion of the last
electron from the quantum dot, providing an absolute reference for
charge counting. By using pulsed gate voltages in combination with
charge sensing, we demonstrate electron tunneling at rates as high as
\amount{2}{MHz} in a one-electron single quantum dot and
\amount{50}{kHz} in a one-electron double quantum dot.

\begin{figure}
\includegraphics[width=9cm]{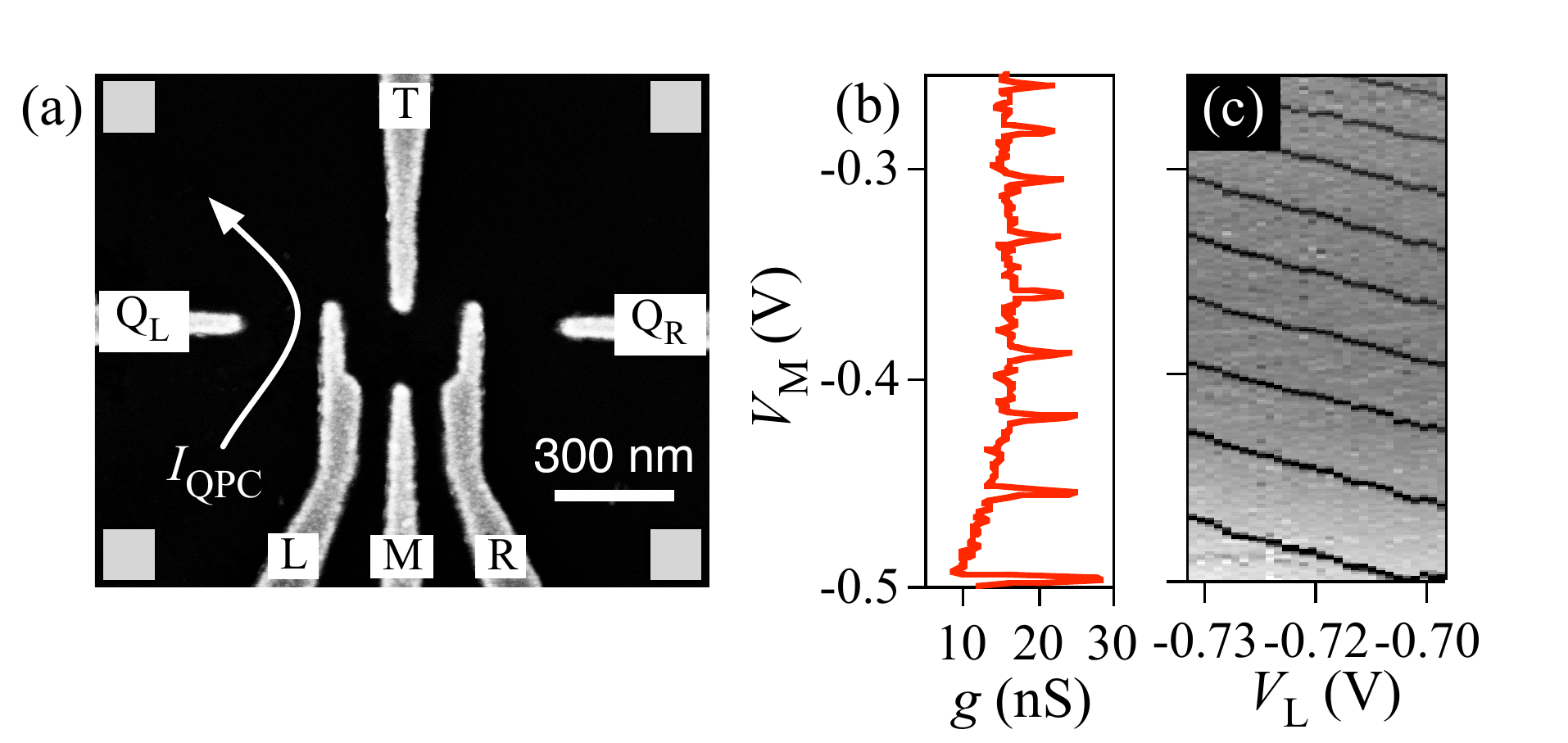}
\caption{\label{fig1}(a) Scanning electron micrograph of the top-gates
  that define the active region of the device. (b) The
  transconductance $g$ of the charge-sensing QPC as a function of
  $V_M$. Peaks in $g$ represent electron tunneling events. (c) Plot of
  $g$ versus the gate voltages $V_M$ and $V_L$. The dark lines
  indicate electron tunneling events and are parallel, showing that
  the device is a single quantum dot in this gate voltage regime.}
\end{figure}

The quantum dot used in this work is formed in a Si/SiGe
heterostructure containing a two-dimensional electron gas
approximately 79 nm below the crystal surface, with a carrier density
of \amount{5.15 \times 10^{11}}{cm^{-2}} and mobility of
\amount{120,000}{cm^{2}/Vs}, after illumination with red light for
\amount{10}{s} while at a temperature of \amount{4.2}{K} at the
beginning of the experiment.  The heterostructure consists of an
undoped relaxed buffer of Si/Si$_{0.71}$Ge$_{0.29}$, a strained-Si
quantum well \amount{18}{nm} thick, \amount{22}{nm} undoped
Si/Si$_{0.71}$Ge$_{0.29}$, \amount{2.6}{nm} phosphrous-doped
Si/Si$_{0.71}$Ge$_{0.29}$, \amount{45}{nm} undoped
Si/Si$_{0.71}$Ge$_{0.29}$, and a \amount{9}{nm} Si cap layer.
Fig.~\ref{fig1}(a) shows a scanning electron micrograph of a device
with the same gate structure as that reported here. The top-gates are
formed by electron-beam evaporation of Pd onto the HF-etched surface
of the heterostructure, following a gate design similar to
Ref.~\cite{Ciorga:2000p16315}. The gates sit on a square mesa of width
\amount{35}{\mu m} that was defined by reactive ion etching.  The
squares at the four corners represent the directions toward available
ohmic contacts.  All measurements reported below were performed in a
dilution refrigerator with a mixing chamber temperature below
\amount{20}{mK}.

The quantum dot is formed by the application of negative voltages to
gates $L$, $M$, $R$ and $T$. The dot is tuned to the few-electron
regime by making the voltages on the gates more negative. Two charge
sensing point contacts are formed by the application of negative
voltage to the left ($Q_L$) and right ($Q_R$) quantum point contact
(QPC) gates; only the data from the left point contact is presented
here. Fig.~\ref{fig1}(b) is a plot of the transconductance $g=\partial
I_{\mathrm{QPC}}/\partial V_M$ as a function of $V_M$, where
$I_{\mathrm{QPC}}$ is the current through the point contact.  The
regularly spaced peaks in $g$ represent changes in the charge
occupation of the dot of one electron. Fig.~\ref{fig1}(c) shows $g$ as
a function of $V_M$ and $V_L$; the peaks in $g$ lie along straight
lines, indicating that the device is a single quantum dot in this gate
voltage regime.

A key question is how to assess when the last observable line in a
plot such as that shown in Fig.~\ref{fig1}(c) represents the removal
of the last electron in the quantum dot.  One typically expects the
changes in gate voltage $\Delta V_G$ to become non-uniform as the last
electron is approached.\cite{Simmons:2007p213103} A concern that is
sometimes raised is whether the tunnel rate into the dot has become
sufficiently slow that an additional transition is simply missed.
This concern is perhaps especially well taken in Si, because the
effective mass of electrons in silicon ($m^*=0.19m_e$) is higher than
that in GaAs ($m^*=0.067m_e$), the most widely studied host for
semiconductor quantum dots. Because of its relatively large $m^*$,
tunnel rates in Si will vary much more strongly as the tunnel barrier
is changed than in GaAs. As a result, care must be taken, in device
design and in experimental measurements, to ensure that the tunnel
barriers do not turn opaque as gate voltages are made more negative
and the few electron regime is reached.  We show that such changes in
tunnel rates yield clear signatures in the data, enabling the device
used here to be reliably tuned to a regime with fast tunnel rates.

\begin{figure}
\includegraphics[width=8cm]{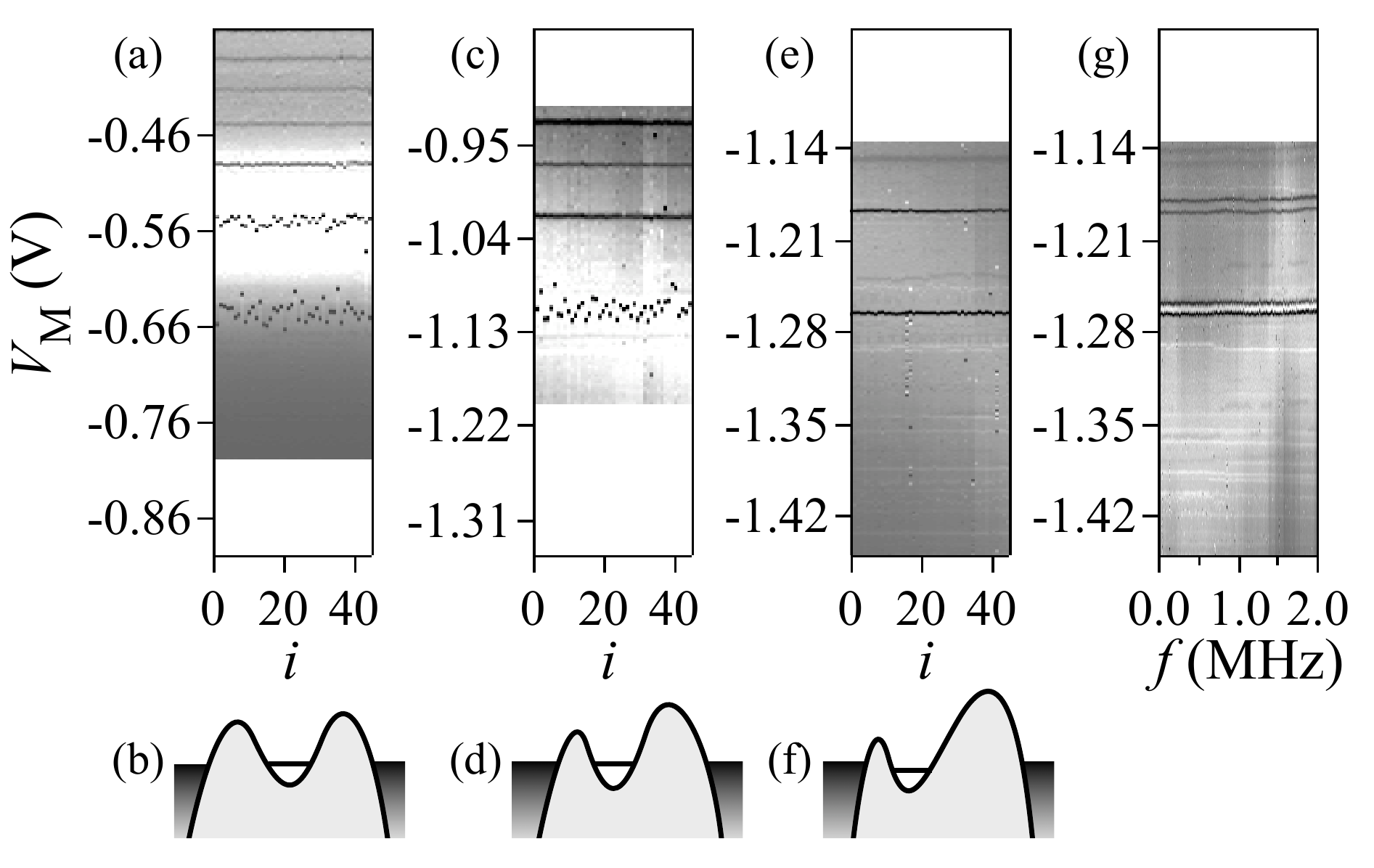}
\caption{\label{fig2} Demonstration of fast and slow electron
  tunneling between the leads and the quantum dot, observable as the
  horizontal black lines. Panels (a), (c), (e) show $g$ as a function
  of $V_M$ for repeated sweeps, labeled with the index $i$. The gate
  voltage conditions are as follows: (a) $V_L$=-0.725 V, $V_R$=-0.83 V
  (c) $V_L$=-0.65 V, $V_R$=-0.94 V (e) $V_L$=-0.54 V, $V_R$=-0.965
  V. The plots are arranged so that the corresponding charge
  transitions are aligned. Panels (b), (d), (f) show schematic
  diagrams of the potential landscape for the plot above each
  diagram. (g) Time-averaged charge sensing measurements in the
  presence of a pulsed gate voltage for the same conditions as panel
  (e): $g$ as a function of $V_M$ and pulse frequency $f$.}
\end{figure}

Fig.~\ref{fig2} shows signatures of both fast and slow tunnel rates.
The gray-scale reports $g$ versus $V_M$ for repeated sweeps of $V_M$
through the same voltage range. For these sweeps, the dwell time per
data point is \amount{250}{ms}, and the voltage step per data point is
\amount{2}{mV} in (a) \& (c) and \amount{1}{mV} in (e) \& (g). The
horizontal axis is an index $i$ representing the sweep number.  Near
the top of Fig.~\ref{fig2}a, the gray lines corresponding to charge
transitions are straight and smooth: electrons hop out of the dot as
soon as the dot chemical potential crosses the Fermi level of the
leads, and the tunneling time $\tau$ is small compared with the dwell
time per data point. As $V_M$ is made more negative, the number of
electrons in the dot decreases, and at the same time cross-talk
between gate $M$ and the quantum dot entrance and exit point contacts
reduces the tunnel coupling between the dot and the leads, increasing
$\tau$.  Eventually $\tau$ approaches and then exceeds the dwell time
per data point, and near the bottom of panel (a) the horizontal lines
are no longer smooth but rather exhibit random fluctuations.  An
important feature of the data in Fig.~\ref{fig2} is that the breaking
up of the charge transition lines is gradual: in Fig.~\ref{fig2}(a),
the line at \amount{V_M = -0.48}{V} is smooth and straight, and the
next two lines, while clearly in the limit of slow $\tau$, are
nonetheless easily discernible.  Fig.~\ref{fig2}(b) shows a schematic
representation of the potential landscape of the dot for the
conditions of panel (a).

Fig.~\ref{fig2}(c) \& (d) show plots analogous to (a) \& (b), with the
voltages $V_L$ and $V_R$ changed to increase the coupling to the left
lead at the expense of that to the right lead.  Because the tunnel
couplings act in parallel, such a procedure reduces the tunneling time
$\tau$, and one expects to see smoother horizontal lines.  In
Fig.~\ref{fig2}(c), the vertical axis is offset to correct for the
effect of changing $V_L$ and $V_R$, lining up the same charge
transitions in (a) and (c).  The second to last charge transition is
now straight and smooth, indicating a faster tunnel rate than in (a),
while the last transition is still uneven.  Note that the sensitivity
of the charge-sensing point contact to electron tunneling events is
also improved in (c) compared with (a), because the changes in gate
voltage shift the position of the dot to the left, closer to the
charge sensing channel.  Fig.~\ref{fig2}(e) \& (f) show the results of
repeating this procedure; again, identical charge transitions are
aligned.  All of the lines are now straight and smooth, no additional
lines have appeared below the final transition, and the sensitivity of
the charge sensing is higher still.

The conditions in Fig.~\ref{fig2}(e) now correspond to tunnel rates
fast enough that relatively high-frequency pulsed-gate voltage
measurements can be performed.  Fig.~\ref{fig2}(g) shows the
time-average of the conductance $g$ in the presence of a square wave
pulse of constant peak-to-peak amplitude \amount{10}{mV} applied to
$V_L$, in addition to its dc voltage. In Fig.~\ref{fig2}(g), every
transition line has split into two lines with no change in intensity
as a function of frequency for frequencies up to \amount{2}{MHz},
indicating that the loading and unloading rates of electrons into and
out of the dot are faster than the pulse frequency over this entire
frequency range.

\begin{figure}
\includegraphics[width=8cm]{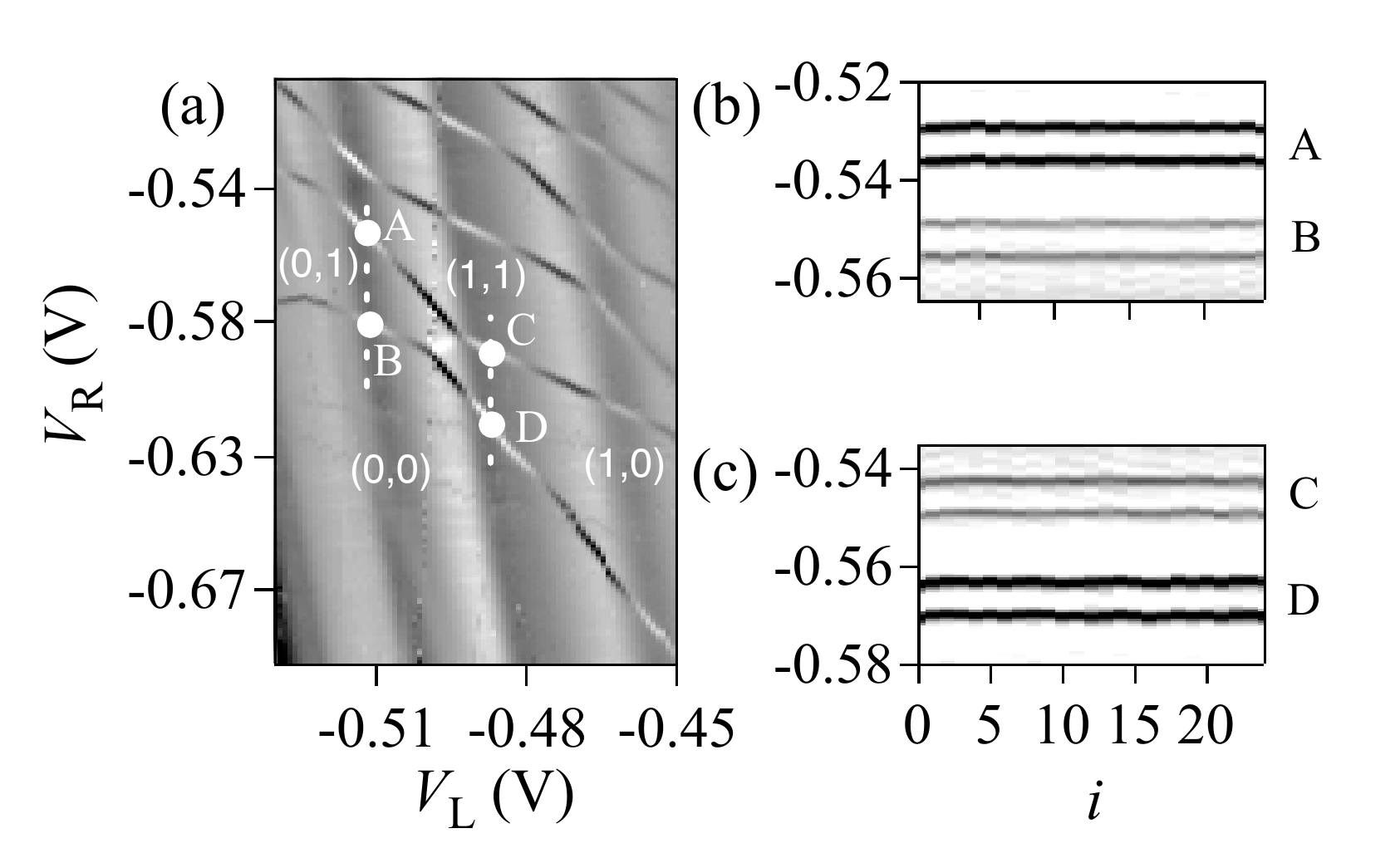}
\caption{\label{fig3}One-electron Si/SiGe double quantum dot. (a)
  Charge stability diagram showing $g$ as a function of $V_L$ and
  $V_R$. (b), (c) Time-averaged pulsed-gate charge sensing
  measurements with a \amount{50}{kHz} square wave of peak-to-peak
  amplitude \amount{7}{mV} added to $V_R$: $g$ is
  plotted as a function of $V_R$ along the range represented by the
  white dashed lines in panel (a).}
\end{figure}

We now reconfigure the voltages on the gates and move smoothly from
the one-electron single dot regime to the one-electron double dot
regime.  To split the single dot into a double dot, $V_T$ is made more
negative, and the other gate voltages are adjusted to keep the one
electron line visible during this procedure.  Fig.~\ref{fig3} (a) is a
plot of $g$ as a function of $V_L$ and $V_R$ in the double dot regime,
showing the characteristic ``honeycomb'' charge transition lines
expected for a double dot. The extended empty region toward the lower
left corner corresponds to the (0,0) charge configuration. We
characterize the tunnel rates of the $(0,0)\rightarrow(0,1)$ and
$(0,1)\rightarrow(1,1)$ transitions by applying voltage pulses to gate
$R$.  Fig.~\ref{fig3}(b) shows $g$ as a function of the sweep index
$i$ and $V_R$, with a \amount{50}{kHz} square-wave pulse of
peak-to-peak amplitude \amount{7}{mV} added to $V_R$.  The white
dashed-line in Fig.~\ref{fig3}(a) represents the sweep taken in $V_R$
for this measurement.  The charge transition lines, corresponding to
the letters A and B in Fig.~\ref{fig3}(a), split into two smooth lines
in Fig.~\ref{fig3}(b), demonstrating that the tunnel rates to access
each of these charge states are at least of the order of
\amount{50}{kHz}, and that the tunneling times $\tau$ are much faster
than the dwell time per data pixel in Fig.~\ref{fig3}(a), which is
\amount{200}{ms}. Fig.~\ref{fig3}(c) shows a similar measurement at
the locations of points C and D.  Note that the data in \ref{fig3}(a)
and that in (b) \& (c) were taken on different days, explaining the
small shift in the operating point between the two sets of data.

We have presented charge sensing data for single and double Si/SiGe
quantum dots in the one-electron regime. We find that tuning the gate
voltages provides good control over the tunnel rates in this regime,
and that pulsed gate voltage measurements can be performed on both
single and double dots. For the single dot, one tunnel barrier can be
made fast by allowing the other tunnel barrier to become slow. Here we
have demonstrated tunnel rates in excess of \amount{2}{MHz} using this
approach. We have also shown that fast tunnel rates can be achieved in
a symmetric double-dot regime, in which both dots are well-coupled to
their corresponding lead. We have demonstrated tunnel rates in excess
of \amount{50}{kHz} to each of those leads. Further increases in
tunnel rates are likely to be achievable for the symmetric operating
regime by changing the size of the lithographic gate pattern.

This work was supported in part by ARO and LPS (W911NF-08-1-0482), by
NSF (DMR-0805045), by United States Department of Defense, and by DOE
(DE-FG02-03ER46028). The views and conclusions contained in this
document are those of the authors and should not be interpreted as
representing the official policies, either expressly or implied, of
the U.S. Government. This research utilized NSF-supported shared
facilities at the University of Wisconsin-Madison.

\bibliography{preprints,silicon,madhu-comments}

\begin{thebibliography}{30}
\expandafter\ifx\csname natexlab\endcsname\relax\def\natexlab#1{#1}\fi
\expandafter\ifx\csname bibnamefont\endcsname\relax
  \def\bibnamefont#1{#1}\fi
\expandafter\ifx\csname bibfnamefont\endcsname\relax
  \def\bibfnamefont#1{#1}\fi
\expandafter\ifx\csname citenamefont\endcsname\relax
  \def\citenamefont#1{#1}\fi
\expandafter\ifx\csname url\endcsname\relax
  \def\url#1{\texttt{#1}}\fi
\expandafter\ifx\csname urlprefix\endcsname\relax\def\urlprefix{URL }\fi
\providecommand{\bibinfo}[2]{#2}
\providecommand{\eprint}[2][]{\url{#2}}

\bibitem[{\citenamefont{Loss and Divincenzo}(1998)}]{Loss:1998p120}
\bibinfo{author}{\bibfnamefont{D.}~\bibnamefont{Loss}} \bibnamefont{and}
  \bibinfo{author}{\bibfnamefont{D.~P.} \bibnamefont{Divincenzo}},
  \bibinfo{journal}{Phys Rev A} \textbf{\bibinfo{volume}{57}},
  \bibinfo{pages}{120} (\bibinfo{year}{1998}).

\bibitem[{\citenamefont{Elzerman et~al.}(2004)\citenamefont{Elzerman, Hanson,
  van Beveren, Witkamp, Vandersypen, and Kouwenhoven}}]{Elzerman:2004p431}
\bibinfo{author}{\bibfnamefont{J.~M.} \bibnamefont{Elzerman}},
  \bibinfo{author}{\bibfnamefont{R.}~\bibnamefont{Hanson}},
  \bibinfo{author}{\bibfnamefont{L.~H.~W.} \bibnamefont{van Beveren}},
  \bibinfo{author}{\bibfnamefont{B.}~\bibnamefont{Witkamp}},
  \bibinfo{author}{\bibfnamefont{L.~M.~K.} \bibnamefont{Vandersypen}},
  \bibnamefont{and} \bibinfo{author}{\bibfnamefont{L.~P.}
  \bibnamefont{Kouwenhoven}}, \bibinfo{journal}{Nature}
  \textbf{\bibinfo{volume}{430}}, \bibinfo{pages}{431} (\bibinfo{year}{2004}).

\bibitem[{\citenamefont{Petta et~al.}(2005)\citenamefont{Petta, Johnson,
  Taylor, Laird, Yacoby, Lukin, Marcus, Hanson, and Gossard}}]{Petta:2005p2180}
\bibinfo{author}{\bibfnamefont{J.~R.} \bibnamefont{Petta}},
  \bibinfo{author}{\bibfnamefont{A.~C.} \bibnamefont{Johnson}},
  \bibinfo{author}{\bibfnamefont{J.~M.} \bibnamefont{Taylor}},
  \bibinfo{author}{\bibfnamefont{E.~A.} \bibnamefont{Laird}},
  \bibinfo{author}{\bibfnamefont{A.}~\bibnamefont{Yacoby}},
  \bibinfo{author}{\bibfnamefont{M.~D.} \bibnamefont{Lukin}},
  \bibinfo{author}{\bibfnamefont{C.~M.} \bibnamefont{Marcus}},
  \bibinfo{author}{\bibfnamefont{M.~P.} \bibnamefont{Hanson}},
  \bibnamefont{and} \bibinfo{author}{\bibfnamefont{A.~C.}
  \bibnamefont{Gossard}}, \bibinfo{journal}{Science}
  \textbf{\bibinfo{volume}{309}}, \bibinfo{pages}{2180} (\bibinfo{year}{2005}).

\bibitem[{\citenamefont{Koppens et~al.}(2006)\citenamefont{Koppens, Buizert,
  Tielrooij, Vink, Nowack, Meunier, Kouwenhoven, and
  Vandersypen}}]{Koppens:2006p766}
\bibinfo{author}{\bibfnamefont{F.~H.~L.} \bibnamefont{Koppens}},
  \bibinfo{author}{\bibfnamefont{C.}~\bibnamefont{Buizert}},
  \bibinfo{author}{\bibfnamefont{K.~J.} \bibnamefont{Tielrooij}},
  \bibinfo{author}{\bibfnamefont{I.~T.} \bibnamefont{Vink}},
  \bibinfo{author}{\bibfnamefont{K.~C.} \bibnamefont{Nowack}},
  \bibinfo{author}{\bibfnamefont{T.}~\bibnamefont{Meunier}},
  \bibinfo{author}{\bibfnamefont{L.~P.} \bibnamefont{Kouwenhoven}},
  \bibnamefont{and} \bibinfo{author}{\bibfnamefont{L.~M.~K.}
  \bibnamefont{Vandersypen}}, \bibinfo{journal}{Nature}
  \textbf{\bibinfo{volume}{442}}, \bibinfo{pages}{766} (\bibinfo{year}{2006}).

\bibitem[{\citenamefont{Tahan et~al.}(2002)\citenamefont{Tahan, Friesen, and
  Joynt}}]{Tahan:2002p035314}
\bibinfo{author}{\bibfnamefont{C.}~\bibnamefont{Tahan}},
  \bibinfo{author}{\bibfnamefont{M.}~\bibnamefont{Friesen}}, \bibnamefont{and}
  \bibinfo{author}{\bibfnamefont{R.}~\bibnamefont{Joynt}},
  \bibinfo{journal}{Phys Rev B} \textbf{\bibinfo{volume}{66}},
  \bibinfo{pages}{035314} (\bibinfo{year}{2002}).

\bibitem[{\citenamefont{Tyryshkin et~al.}(2003)\citenamefont{Tyryshkin, Lyon,
  Astashkin, and Raitsimring}}]{Tyryshkin:2003p193207}
\bibinfo{author}{\bibfnamefont{A.~M.} \bibnamefont{Tyryshkin}},
  \bibinfo{author}{\bibfnamefont{S.~A.} \bibnamefont{Lyon}},
  \bibinfo{author}{\bibfnamefont{A.~V.} \bibnamefont{Astashkin}},
  \bibnamefont{and} \bibinfo{author}{\bibfnamefont{A.~M.}
  \bibnamefont{Raitsimring}}, \bibinfo{journal}{Phys Rev B}
  \textbf{\bibinfo{volume}{68}}, \bibinfo{pages}{193207}
  (\bibinfo{year}{2003}).

\bibitem[{\citenamefont{de~Sousa and Das~Sarma}(2003)}]{deSousa:2003p1660}
\bibinfo{author}{\bibfnamefont{R.}~\bibnamefont{de~Sousa}} \bibnamefont{and}
  \bibinfo{author}{\bibfnamefont{S.}~\bibnamefont{Das~Sarma}},
  \bibinfo{journal}{Phys Rev B} \textbf{\bibinfo{volume}{68}},
  \bibinfo{pages}{115322} (\bibinfo{year}{2003}).

\bibitem[{\citenamefont{Kane}(1998)}]{Kane:1998p133}
\bibinfo{author}{\bibfnamefont{B.~E.} \bibnamefont{Kane}},
  \bibinfo{journal}{Nature} \textbf{\bibinfo{volume}{393}},
  \bibinfo{pages}{133} (\bibinfo{year}{1998}).

\bibitem[{\citenamefont{Vrijen et~al.}(2000)\citenamefont{Vrijen, Yablonovitch,
  Wang, Jiang, Balandin, Roychowdhury, Mor, and Divincenzo}}]{Vrijen:2000p1643}
\bibinfo{author}{\bibfnamefont{R.}~\bibnamefont{Vrijen}},
  \bibinfo{author}{\bibfnamefont{E.}~\bibnamefont{Yablonovitch}},
  \bibinfo{author}{\bibfnamefont{K.}~\bibnamefont{Wang}},
  \bibinfo{author}{\bibfnamefont{H.~W.} \bibnamefont{Jiang}},
  \bibinfo{author}{\bibfnamefont{A.}~\bibnamefont{Balandin}},
  \bibinfo{author}{\bibfnamefont{V.}~\bibnamefont{Roychowdhury}},
  \bibinfo{author}{\bibfnamefont{T.}~\bibnamefont{Mor}}, \bibnamefont{and}
  \bibinfo{author}{\bibfnamefont{D.}~\bibnamefont{Divincenzo}},
  \bibinfo{journal}{Phys. Rev. A} \textbf{\bibinfo{volume}{62}},
  \bibinfo{pages}{012306} (\bibinfo{year}{2000}).

\bibitem[{\citenamefont{Friesen et~al.}(2003)\citenamefont{Friesen, Rugheimer,
  Savage, Lagally, van~der Weide, Joynt, and Eriksson}}]{Friesen:2003p121301}
\bibinfo{author}{\bibfnamefont{M.}~\bibnamefont{Friesen}},
  \bibinfo{author}{\bibfnamefont{P.}~\bibnamefont{Rugheimer}},
  \bibinfo{author}{\bibfnamefont{D.~E.} \bibnamefont{Savage}},
  \bibinfo{author}{\bibfnamefont{M.~G.} \bibnamefont{Lagally}},
  \bibinfo{author}{\bibfnamefont{D.~W.} \bibnamefont{van~der Weide}},
  \bibinfo{author}{\bibfnamefont{R.}~\bibnamefont{Joynt}}, \bibnamefont{and}
  \bibinfo{author}{\bibfnamefont{M.~A.} \bibnamefont{Eriksson}},
  \bibinfo{journal}{Phys Rev B} \textbf{\bibinfo{volume}{67}},
  \bibinfo{pages}{121301} (\bibinfo{year}{2003}).

\bibitem[{\citenamefont{Hollenberg et~al.}(2006)\citenamefont{Hollenberg,
  Greentree, Fowler, and Wellard}}]{Hollenberg:2006p1534}
\bibinfo{author}{\bibfnamefont{L.~C.~L.} \bibnamefont{Hollenberg}},
  \bibinfo{author}{\bibfnamefont{A.~D.} \bibnamefont{Greentree}},
  \bibinfo{author}{\bibfnamefont{A.~G.} \bibnamefont{Fowler}},
  \bibnamefont{and} \bibinfo{author}{\bibfnamefont{C.~J.}
  \bibnamefont{Wellard}}, \bibinfo{journal}{Phys Rev B}
  \textbf{\bibinfo{volume}{74}}, \bibinfo{pages}{045311}
  (\bibinfo{year}{2006}).

\bibitem[{\citenamefont{Morello et~al.}(2009)\citenamefont{Morello, Escott,
  Huebl, van Beveren, Hollenberg, Jamieson, Dzurak, and
  Clark}}]{Morello:2009p081307}
\bibinfo{author}{\bibfnamefont{A.}~\bibnamefont{Morello}},
  \bibinfo{author}{\bibfnamefont{C.~C.} \bibnamefont{Escott}},
  \bibinfo{author}{\bibfnamefont{H.}~\bibnamefont{Huebl}},
  \bibinfo{author}{\bibfnamefont{L.~H.~W.} \bibnamefont{van Beveren}},
  \bibinfo{author}{\bibfnamefont{L.~C.~L.} \bibnamefont{Hollenberg}},
  \bibinfo{author}{\bibfnamefont{D.~N.} \bibnamefont{Jamieson}},
  \bibinfo{author}{\bibfnamefont{A.~S.} \bibnamefont{Dzurak}},
  \bibnamefont{and} \bibinfo{author}{\bibfnamefont{R.~G.} \bibnamefont{Clark}},
  \bibinfo{journal}{Phys Rev B} \textbf{\bibinfo{volume}{80}},
  \bibinfo{pages}{081307} (\bibinfo{year}{2009}).

\bibitem[{\citenamefont{Slinker et~al.}(2005)\citenamefont{Slinker, Lewis,
  Haselby, Goswami, Klein, Chu, Coppersmith, Joynt, Blick, Friesen
  et~al.}}]{Slinker:2005p246}
\bibinfo{author}{\bibfnamefont{K.~A.} \bibnamefont{Slinker}},
  \bibinfo{author}{\bibfnamefont{K.~L.~M.} \bibnamefont{Lewis}},
  \bibinfo{author}{\bibfnamefont{C.~C.} \bibnamefont{Haselby}},
  \bibinfo{author}{\bibfnamefont{S.}~\bibnamefont{Goswami}},
  \bibinfo{author}{\bibfnamefont{L.~J.} \bibnamefont{Klein}},
  \bibinfo{author}{\bibfnamefont{J.~O.} \bibnamefont{Chu}},
  \bibinfo{author}{\bibfnamefont{S.~N.} \bibnamefont{Coppersmith}},
  \bibinfo{author}{\bibfnamefont{R.}~\bibnamefont{Joynt}},
  \bibinfo{author}{\bibfnamefont{R.~H.} \bibnamefont{Blick}},
  \bibinfo{author}{\bibfnamefont{M.}~\bibnamefont{Friesen}},
  \bibnamefont{et~al.}, \bibinfo{journal}{New J Phys}
  \textbf{\bibinfo{volume}{7}}, \bibinfo{pages}{246} (\bibinfo{year}{2005}).

\bibitem[{\citenamefont{Berer et~al.}(2006)\citenamefont{Berer, Pachinger,
  Pillwein, M\"{u}hlberger, Lichtenberger, Brunthaler, and
  Sch\"{a}ffler}}]{Berer:2006p162112}
\bibinfo{author}{\bibfnamefont{T.}~\bibnamefont{Berer}},
  \bibinfo{author}{\bibfnamefont{D.}~\bibnamefont{Pachinger}},
  \bibinfo{author}{\bibfnamefont{G.}~\bibnamefont{Pillwein}},
  \bibinfo{author}{\bibfnamefont{M.}~\bibnamefont{M\"{u}hlberger}},
  \bibinfo{author}{\bibfnamefont{H.}~\bibnamefont{Lichtenberger}},
  \bibinfo{author}{\bibfnamefont{G.}~\bibnamefont{Brunthaler}},
  \bibnamefont{and}
  \bibinfo{author}{\bibfnamefont{F.}~\bibnamefont{Sch\"{a}ffler}},
  \bibinfo{journal}{Appl Phys Lett} \textbf{\bibinfo{volume}{88}},
  \bibinfo{pages}{162112} (\bibinfo{year}{2006}).

\bibitem[{\citenamefont{Angus et~al.}(2007)\citenamefont{Angus, Ferguson,
  Dzurak, and Clark}}]{Angus:2007p845}
\bibinfo{author}{\bibfnamefont{S.~J.} \bibnamefont{Angus}},
  \bibinfo{author}{\bibfnamefont{A.~J.} \bibnamefont{Ferguson}},
  \bibinfo{author}{\bibfnamefont{A.~S.} \bibnamefont{Dzurak}},
  \bibnamefont{and} \bibinfo{author}{\bibfnamefont{R.~G.} \bibnamefont{Clark}},
  \bibinfo{journal}{Nano Lett} \textbf{\bibinfo{volume}{7}},
  \bibinfo{pages}{2051} (\bibinfo{year}{2007}).

\bibitem[{\citenamefont{Hu et~al.}(2007)\citenamefont{Hu, Churchill, Reilly,
  Xiang, Lieber, and Marcus}}]{Hu:2007p465}
\bibinfo{author}{\bibfnamefont{Y.}~\bibnamefont{Hu}},
  \bibinfo{author}{\bibfnamefont{H.~O.~H.} \bibnamefont{Churchill}},
  \bibinfo{author}{\bibfnamefont{D.~J.} \bibnamefont{Reilly}},
  \bibinfo{author}{\bibfnamefont{J.}~\bibnamefont{Xiang}},
  \bibinfo{author}{\bibfnamefont{C.~M.} \bibnamefont{Lieber}},
  \bibnamefont{and} \bibinfo{author}{\bibfnamefont{C.~M.}
  \bibnamefont{Marcus}}, \bibinfo{journal}{Nat Nanotechnol}
  \textbf{\bibinfo{volume}{2}}, \bibinfo{pages}{622} (\bibinfo{year}{2007}).

\bibitem[{\citenamefont{Zimmerman et~al.}(2007)\citenamefont{Zimmerman,
  Simonds, Fujiwara, Ono, Takahashi, and Inokawa}}]{Zimmerman:2007p033507}
\bibinfo{author}{\bibfnamefont{N.~M.} \bibnamefont{Zimmerman}},
  \bibinfo{author}{\bibfnamefont{B.~J.} \bibnamefont{Simonds}},
  \bibinfo{author}{\bibfnamefont{A.}~\bibnamefont{Fujiwara}},
  \bibinfo{author}{\bibfnamefont{Y.}~\bibnamefont{Ono}},
  \bibinfo{author}{\bibfnamefont{Y.}~\bibnamefont{Takahashi}},
  \bibnamefont{and} \bibinfo{author}{\bibfnamefont{H.}~\bibnamefont{Inokawa}},
  \bibinfo{journal}{Appl Phys Lett} \textbf{\bibinfo{volume}{90}},
  \bibinfo{pages}{033507} (\bibinfo{year}{2007}).

\bibitem[{\citenamefont{Shaji et~al.}(2008)\citenamefont{Shaji, Simmons,
  Thalakulam, Klein, Qin, Luo, Savage, Lagally, Rimberg, Joynt
  et~al.}}]{Shaji:2008p540}
\bibinfo{author}{\bibfnamefont{N.}~\bibnamefont{Shaji}},
  \bibinfo{author}{\bibfnamefont{C.~B.} \bibnamefont{Simmons}},
  \bibinfo{author}{\bibfnamefont{M.}~\bibnamefont{Thalakulam}},
  \bibinfo{author}{\bibfnamefont{L.~J.} \bibnamefont{Klein}},
  \bibinfo{author}{\bibfnamefont{H.}~\bibnamefont{Qin}},
  \bibinfo{author}{\bibfnamefont{H.}~\bibnamefont{Luo}},
  \bibinfo{author}{\bibfnamefont{D.~E.} \bibnamefont{Savage}},
  \bibinfo{author}{\bibfnamefont{M.~G.} \bibnamefont{Lagally}},
  \bibinfo{author}{\bibfnamefont{A.~J.} \bibnamefont{Rimberg}},
  \bibinfo{author}{\bibfnamefont{R.}~\bibnamefont{Joynt}},
  \bibnamefont{et~al.}, \bibinfo{journal}{Nat Phys}
  \textbf{\bibinfo{volume}{4}}, \bibinfo{pages}{540} (\bibinfo{year}{2008}).

\bibitem[{\citenamefont{Liu et~al.}(2008)\citenamefont{Liu, Fujisawa, Ono,
  Inokawa, Fujiwara, Takashina, and Hirayama}}]{Liu:2008p073310}
\bibinfo{author}{\bibfnamefont{H.~W.} \bibnamefont{Liu}},
  \bibinfo{author}{\bibfnamefont{T.}~\bibnamefont{Fujisawa}},
  \bibinfo{author}{\bibfnamefont{Y.}~\bibnamefont{Ono}},
  \bibinfo{author}{\bibfnamefont{H.}~\bibnamefont{Inokawa}},
  \bibinfo{author}{\bibfnamefont{A.}~\bibnamefont{Fujiwara}},
  \bibinfo{author}{\bibfnamefont{K.}~\bibnamefont{Takashina}},
  \bibnamefont{and} \bibinfo{author}{\bibfnamefont{Y.}~\bibnamefont{Hirayama}},
  \bibinfo{journal}{Phys Rev B} \textbf{\bibinfo{volume}{77}},
  \bibinfo{pages}{073310} (\bibinfo{year}{2008}).

\bibitem[{\citenamefont{Lansbergen et~al.}(2008)\citenamefont{Lansbergen,
  Rahman, Wellard, Woo, Caro, Collaert, Biesemans, Klimeck, Hollenberg, and
  Rogge}}]{Lansbergen:2008p1545}
\bibinfo{author}{\bibfnamefont{G.~P.} \bibnamefont{Lansbergen}},
  \bibinfo{author}{\bibfnamefont{R.}~\bibnamefont{Rahman}},
  \bibinfo{author}{\bibfnamefont{C.~J.} \bibnamefont{Wellard}},
  \bibinfo{author}{\bibfnamefont{I.}~\bibnamefont{Woo}},
  \bibinfo{author}{\bibfnamefont{J.}~\bibnamefont{Caro}},
  \bibinfo{author}{\bibfnamefont{N.}~\bibnamefont{Collaert}},
  \bibinfo{author}{\bibfnamefont{S.}~\bibnamefont{Biesemans}},
  \bibinfo{author}{\bibfnamefont{G.}~\bibnamefont{Klimeck}},
  \bibinfo{author}{\bibfnamefont{L.~C.~L.} \bibnamefont{Hollenberg}},
  \bibnamefont{and} \bibinfo{author}{\bibfnamefont{S.}~\bibnamefont{Rogge}},
  \bibinfo{journal}{Nat Phys} \textbf{\bibinfo{volume}{4}},
  \bibinfo{pages}{656} (\bibinfo{year}{2008}).

\bibitem[{\citenamefont{Fuhrer et~al.}(2009)\citenamefont{Fuhrer, F\"uchsle,
  Reusch, Weber, and Simmons}}]{Fuhrer:2009p707}
\bibinfo{author}{\bibfnamefont{A.}~\bibnamefont{Fuhrer}},
  \bibinfo{author}{\bibfnamefont{M.}~\bibnamefont{F\"uchsle}},
  \bibinfo{author}{\bibfnamefont{T.~C.~G.} \bibnamefont{Reusch}},
  \bibinfo{author}{\bibfnamefont{B.}~\bibnamefont{Weber}}, \bibnamefont{and}
  \bibinfo{author}{\bibfnamefont{M.~Y.} \bibnamefont{Simmons}},
  \bibinfo{journal}{Nano Lett} \textbf{\bibinfo{volume}{9}},
  \bibinfo{pages}{707} (\bibinfo{year}{2009}).

\bibitem[{\citenamefont{Simmons et~al.}(2009)\citenamefont{Simmons, Thalakulam,
  Rosemeyer, Bael, Sackmann, Savage, Lagally, Joynt, Friesen, Coppersmith
  et~al.}}]{Simmons:2009p3234}
\bibinfo{author}{\bibfnamefont{C.~B.} \bibnamefont{Simmons}},
  \bibinfo{author}{\bibfnamefont{M.}~\bibnamefont{Thalakulam}},
  \bibinfo{author}{\bibfnamefont{B.~M.} \bibnamefont{Rosemeyer}},
  \bibinfo{author}{\bibfnamefont{B.~J.~V.} \bibnamefont{Bael}},
  \bibinfo{author}{\bibfnamefont{E.~K.} \bibnamefont{Sackmann}},
  \bibinfo{author}{\bibfnamefont{D.~E.} \bibnamefont{Savage}},
  \bibinfo{author}{\bibfnamefont{M.~G.} \bibnamefont{Lagally}},
  \bibinfo{author}{\bibfnamefont{R.}~\bibnamefont{Joynt}},
  \bibinfo{author}{\bibfnamefont{M.}~\bibnamefont{Friesen}},
  \bibinfo{author}{\bibfnamefont{S.~N.} \bibnamefont{Coppersmith}},
  \bibnamefont{et~al.}, \bibinfo{journal}{Nano Lett}
  \textbf{\bibinfo{volume}{9}}, \bibinfo{pages}{3234} (\bibinfo{year}{2009}).

\bibitem[{\citenamefont{Nordberg et~al.}(2009)\citenamefont{Nordberg, Eyck,
  Stalford, Muller, Young, Eng, Tracy, Childs, Wendt, Grubbs
  et~al.}}]{Nordberg:2009p115331}
\bibinfo{author}{\bibfnamefont{E.~P.} \bibnamefont{Nordberg}},
  \bibinfo{author}{\bibfnamefont{G.~A.~T.} \bibnamefont{Eyck}},
  \bibinfo{author}{\bibfnamefont{H.~L.} \bibnamefont{Stalford}},
  \bibinfo{author}{\bibfnamefont{R.~P.} \bibnamefont{Muller}},
  \bibinfo{author}{\bibfnamefont{R.~W.} \bibnamefont{Young}},
  \bibinfo{author}{\bibfnamefont{K.}~\bibnamefont{Eng}},
  \bibinfo{author}{\bibfnamefont{L.~A.} \bibnamefont{Tracy}},
  \bibinfo{author}{\bibfnamefont{K.~D.} \bibnamefont{Childs}},
  \bibinfo{author}{\bibfnamefont{J.~R.} \bibnamefont{Wendt}},
  \bibinfo{author}{\bibfnamefont{R.~K.} \bibnamefont{Grubbs}},
  \bibnamefont{et~al.}, \bibinfo{journal}{Phys Rev B}
  \textbf{\bibinfo{volume}{80}}, \bibinfo{pages}{115331}
  (\bibinfo{year}{2009}).

\bibitem[{\citenamefont{Hayes et~al.}(2009)\citenamefont{Hayes, Kiselev,
  Borselli, Bui, III, Deelman, Maune, Milosavljevic, Moon, Ross
  et~al.}}]{Hayes:2009preprint}
\bibinfo{author}{\bibfnamefont{R.~R.} \bibnamefont{Hayes}},
  \bibinfo{author}{\bibfnamefont{A.~A.} \bibnamefont{Kiselev}},
  \bibinfo{author}{\bibfnamefont{M.~G.} \bibnamefont{Borselli}},
  \bibinfo{author}{\bibfnamefont{S.~S.} \bibnamefont{Bui}},
  \bibinfo{author}{\bibfnamefont{E.~T.~C.} \bibnamefont{III}},
  \bibinfo{author}{\bibfnamefont{P.~W.} \bibnamefont{Deelman}},
  \bibinfo{author}{\bibfnamefont{B.~M.} \bibnamefont{Maune}},
  \bibinfo{author}{\bibfnamefont{I.}~\bibnamefont{Milosavljevic}},
  \bibinfo{author}{\bibfnamefont{J.-S.} \bibnamefont{Moon}},
  \bibinfo{author}{\bibfnamefont{R.~S.} \bibnamefont{Ross}},
  \bibnamefont{et~al.} (\bibinfo{year}{2009}), \bibinfo{note}{arXiv:0908.0173}.

\bibitem[{\citenamefont{Xiao et~al.}(2009)\citenamefont{Xiao, House, and
  Jiang}}]{Xiao:2009preprint}
\bibinfo{author}{\bibfnamefont{M.}~\bibnamefont{Xiao}},
  \bibinfo{author}{\bibfnamefont{M.~G.} \bibnamefont{House}}, \bibnamefont{and}
  \bibinfo{author}{\bibfnamefont{H.~W.} \bibnamefont{Jiang}}
  (\bibinfo{year}{2009}), \bibinfo{note}{arXiv:0909.2857v1}.

\bibitem[{\citenamefont{Simmons et~al.}(2007)\citenamefont{Simmons, Thalakulam,
  Shaji, Klein, Qin, Blick, Savage, Lagally, Coppersmith, and
  Eriksson}}]{Simmons:2007p213103}
\bibinfo{author}{\bibfnamefont{C.~B.} \bibnamefont{Simmons}},
  \bibinfo{author}{\bibfnamefont{M.}~\bibnamefont{Thalakulam}},
  \bibinfo{author}{\bibfnamefont{N.}~\bibnamefont{Shaji}},
  \bibinfo{author}{\bibfnamefont{L.~J.} \bibnamefont{Klein}},
  \bibinfo{author}{\bibfnamefont{H.}~\bibnamefont{Qin}},
  \bibinfo{author}{\bibfnamefont{R.~H.} \bibnamefont{Blick}},
  \bibinfo{author}{\bibfnamefont{D.~E.} \bibnamefont{Savage}},
  \bibinfo{author}{\bibfnamefont{M.~G.} \bibnamefont{Lagally}},
  \bibinfo{author}{\bibfnamefont{S.~N.} \bibnamefont{Coppersmith}},
  \bibnamefont{and} \bibinfo{author}{\bibfnamefont{M.~A.}
  \bibnamefont{Eriksson}}, \bibinfo{journal}{Appl Phys Lett}
  \textbf{\bibinfo{volume}{91}}, \bibinfo{pages}{213103}
  (\bibinfo{year}{2007}).

\bibitem[{\citenamefont{Lim et~al.}(2009)\citenamefont{Lim, Zwanenburg, Huebl,
  M{\"o}tt{\"o}nen, Chan, Morello, and Dzurak}}]{Lim:2009p242102}
\bibinfo{author}{\bibfnamefont{W.~H.} \bibnamefont{Lim}},
  \bibinfo{author}{\bibfnamefont{F.~A.} \bibnamefont{Zwanenburg}},
  \bibinfo{author}{\bibfnamefont{H.}~\bibnamefont{Huebl}},
  \bibinfo{author}{\bibfnamefont{M.}~\bibnamefont{M{\"o}tt{\"o}nen}},
  \bibinfo{author}{\bibfnamefont{K.~W.} \bibnamefont{Chan}},
  \bibinfo{author}{\bibfnamefont{A.}~\bibnamefont{Morello}}, \bibnamefont{and}
  \bibinfo{author}{\bibfnamefont{A.~S.} \bibnamefont{Dzurak}},
  \bibinfo{journal}{Appl Phys Lett} \textbf{\bibinfo{volume}{95}},
  \bibinfo{pages}{242102} (\bibinfo{year}{2009}).

\bibitem[{\citenamefont{Zwanenburg et~al.}(2009)\citenamefont{Zwanenburg,
  Rijmenam, Fang, Lieber, and Kouwenhoven}}]{Zwanenburg:2009p1071}
\bibinfo{author}{\bibfnamefont{F.~A.} \bibnamefont{Zwanenburg}},
  \bibinfo{author}{\bibfnamefont{C.~E. W. M.~V.} \bibnamefont{Rijmenam}},
  \bibinfo{author}{\bibfnamefont{Y.}~\bibnamefont{Fang}},
  \bibinfo{author}{\bibfnamefont{C.~M.} \bibnamefont{Lieber}},
  \bibnamefont{and} \bibinfo{author}{\bibfnamefont{L.~P.}
  \bibnamefont{Kouwenhoven}}, \bibinfo{journal}{Nano Lett}
  \textbf{\bibinfo{volume}{9}}, \bibinfo{pages}{1071} (\bibinfo{year}{2009}).

\bibitem[{\citenamefont{Taylor et~al.}(2005)\citenamefont{Taylor, Engel, Dur,
  Yacoby, Marcus, Zoller, and Lukin}}]{Taylor:2005p482}
\bibinfo{author}{\bibfnamefont{J.}~\bibnamefont{Taylor}},
  \bibinfo{author}{\bibfnamefont{H.}~\bibnamefont{Engel}},
  \bibinfo{author}{\bibfnamefont{W.}~\bibnamefont{Dur}},
  \bibinfo{author}{\bibfnamefont{A.}~\bibnamefont{Yacoby}},
  \bibinfo{author}{\bibfnamefont{C.}~\bibnamefont{Marcus}},
  \bibinfo{author}{\bibfnamefont{P.}~\bibnamefont{Zoller}}, \bibnamefont{and}
  \bibinfo{author}{\bibfnamefont{M.}~\bibnamefont{Lukin}},
  \bibinfo{journal}{Nat Phys} \textbf{\bibinfo{volume}{1}},
  \bibinfo{pages}{177} (\bibinfo{year}{2005}).

\bibitem[{\citenamefont{Ciorga et~al.}(2000)\citenamefont{Ciorga, Sachrajda,
  Hawrylak, Gould, Zawadzki, Jullian, Feng, and
  Wasilewski}}]{Ciorga:2000p16315}
\bibinfo{author}{\bibfnamefont{M.}~\bibnamefont{Ciorga}},
  \bibinfo{author}{\bibfnamefont{A.~S.} \bibnamefont{Sachrajda}},
  \bibinfo{author}{\bibfnamefont{P.}~\bibnamefont{Hawrylak}},
  \bibinfo{author}{\bibfnamefont{C.}~\bibnamefont{Gould}},
  \bibinfo{author}{\bibfnamefont{P.}~\bibnamefont{Zawadzki}},
  \bibinfo{author}{\bibfnamefont{S.}~\bibnamefont{Jullian}},
  \bibinfo{author}{\bibfnamefont{Y.}~\bibnamefont{Feng}}, \bibnamefont{and}
  \bibinfo{author}{\bibfnamefont{Z.}~\bibnamefont{Wasilewski}},
  \bibinfo{journal}{Phys Rev B} \textbf{\bibinfo{volume}{61}},
  \bibinfo{pages}{R16315} (\bibinfo{year}{2000}).

\end{thebibliography}

\end{document}